\documentstyle[sprocl,epsf]{article}

\newcommand{\cF}{{\cal F}}
\newcommand{\cA}{{\cal A}}

\newcommand{\cM}{{\cal M}}

\newcommand{\bi}{\bigskip}
\newcommand{\no}{\noindent}
\newcommand{\be}{\begin{eqnarray}}
\newcommand{\ee}{\end{eqnarray}}
\newcommand{\hk}{\hspace{0.1cm}}

\newcommand{\rk}{\right)}
\newcommand{\lk}{\left(}

\newcommand{\sli}{\sum\limits}

\newcommand{\il}{\int\limits}

\def\be{\begin{equation}}
\def\ee{\end{equation}}
\def\bea{\begin{eqnarray}}
\def\eea{\end{eqnarray}}

\begin{document}

\title{CENTER VORTICES IN CONTINUUM YANG-MILLS THEORY\footnote{Invited
talk presented by H. Reinhardt at the International Conference on 
``Confinement and the Structure of Hadrons'' in Vienna, 3.-8.7.00.}}

\author{H. REINHARDT\footnote{Supported in part by DFG Re 856/4-1} and M. 
ENGELHARDT\footnote{Supported by DFG under En 415/1-1.} }

\address{Institut f\"ur Theoretische Physik, Universit\"at T\"ubingen\\
Auf der Morgenstelle 14, 72076 T\"ubingen, Germany\\
E-mail: hugo.reinhardt@uni-tuebingen.de} 



\maketitle\abstracts{The properties of center vortices are discussed within
continuum Yang-Mills theory. By starting from the lattice theory and carefully
performing the continuum limit the gauge potential of center vortices is
obtained and the continuum analog of the maximal center gauge fixing is
extracted. It is shown, that the Pontryagin index of center vortices is given by
their self-intersection number, which vanishes unless the center vortices host
magnetic monpoles, which make the vortex sheets non-oriented. }

\section{Introduction}
\bi

\no
At present there are two popular confinement mechanisms: the dual Meissner 
effect, which is based on a condensation of magnetic monopoles in the QCD 
vacuum, and the vortex condensation picture. Both pictures were proposed 
a long time ago, but only in recent years mounting evidence for the 
realization of these pictures has been accumulated in lattice calculations. 
Both pictures show up in specific partial gauge fixings.
\smallskip

\no
The vortex picture of confinement, which received rather little attention after
some early efforts following its inception, has recently received
strong support from lattice calculations performed in the so-called maximal
center gauge, where one fixes only the coset $G / Z$, but leaves the center $Z$
of the gauge group $G$ unfixed \cite{R1}. Subsequent center projection, which 
consists of
replacing each link by its closest center element, allows the identification of
the center vortex content of the gauge fields. It is found that the vortex
content is a physical property of the gauge ensemble \cite{R2} and
produces virtually the full Yang-Mills string tension \cite{R1}.
Conversely the string tension disappears in the absence of center vortices. 
This property of center dominance persists at finite temperature \cite{R3} 
and the deconfinement phase transition
can be understood in a 3-dimensional slice at a fixed spatial coordinate 
as a transition from a percolating vortex phase to a phase in
which vortices cease to percolate \cite{R3,selprep}.
The vortices have also been shown to condense
in the confinement phase \cite{R4}. Furthermore, in a gauge field
ensemble devoid of center vortices, chiral symmetry breaking disappears and all
field configurations belong to the topologically trivial sector \cite{R5}. 
In fact, the
magnetic degrees of freedom causing confinement, monopoles and vortices, can be
connected with the topological properties commonly thought to be carried by
instantons. 
\bi

\no
In this lecture, I will discuss center vortices from the point of view of
continuum Yang-Mills theory and study their topological properties. We will
find that center vortices with non-vanishing Pontryagin index carry magnetic
monopole loops. The talk is largely based on ref. \cite{R6}.

\section{Center vortices in continuum Yang-Mills theory}
\bi

\no
As a warming-up exercise I review how center vortices are identified in lattice
gauge theory by center projection after implementing the maximal center gauge.
\bi

\no
The lattice gauge theory is defined in terms of link variables $U_\mu (x) \in
G$. In the maximal center
gauge the gauge freedom of these link variables is exploited 
to bring the links $U_\mu (x) \in G$ as close as possible to a center element
of the gauge group,
\be
\label{n3}
\sli_{x, \nu} \lk tr U_\mu (x) \rk^2 \to max. \hk 
\ee
For simplicity let us consider the gauge group $SU (2)$. The
center is given by $Z (2) = \{1, -1 \}$. 
The condition (\ref{n3}) 
is insensitive to a change of the sign of the link variable $U_\mu \to (-
U_\mu)$, which just
reflects the residual $Z (2)$ symmetry. Once the maximal center gauge is
implemented, center projection implies to put by hand
$U_\mu (x) \to sign \lk tr U_\mu (x) \rk$.
After center projection, all links become center elements, that is for 
$SU (2)$, $U_\mu (x) = \pm 1$ .
As a result of the center projection $D - 1$ dimensional domains (hypersurfaces)
$\Sigma$ of links equal to a non-trivial center element arise in a background of
links given by $U_\mu (x) = 1$, see fig. 1. 
\begin{figure}
\centerline{
\epsfysize=5cm
\epsffile{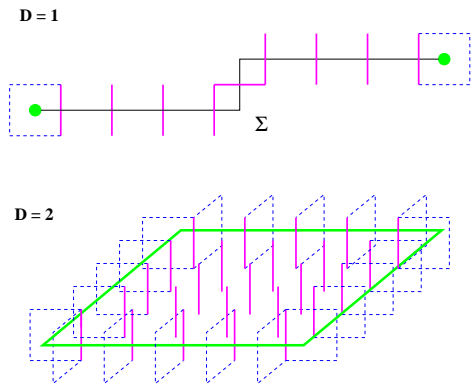}
}
\caption{Illustration of non-trivial center vortex configurations arising after
center projection in two and three dimensions. The full links represent
non-trivial center elements forming the hypersurface $\Sigma$, the boundary
$\partial \Sigma$ of which represents the center vortex}
\label{fig1}
\end{figure}

The boundaries of these
hypersurfaces are given by closed $D - 2$ dimensional surfaces $\partial
\Sigma$, which live on the dual lattice and define an ideal center vortex. When
the boundary $\partial \Sigma$ is non-trivially linked to a Wilson loop, the
latter receives the corresponding non-trivial center element $Z = - 1$. 
This can be
also seen in the following way. As one travels along the loop, accumulating
phases from different links making up the path, the Wilson loop stays either
constant or picks up a phase corresponding to a non-trivial element of the
center of the gauge group on a length of a lattice spacing. This latter change
happens whenever the Wilson loops intersect the $D - 1$ dimensional hypersurface
$\Sigma$ 
describing an ideal center vortex configuration $\partial \Sigma$.
\smallskip

\no
The concept of introducing center vortices by center projection can be
straightforwardly extended to higher gauge groups $SU (N > 2)$, for which the
center elements are given by the N'th roots of unity
$Z (k) = e^{i \frac{2 \pi}{N} k} \hk , \hk k = 0, 1, \dots, N - 1$.
Vortices are still given by the boundaries of $D - 1$ domains of links
equal to a definite non-trivial center element $Z (k)$. Since for $N > 2$
multiplication of non-trivial center elements can yield another non-trivial
center element the possibility of center vortex fusion and fission exists. For
example, for $SU (3)$ the center consists of the set of elements
$\left\{ z = e^{i \frac{2}{3} \pi} \hk , \hk z^2 = e^{i \frac{4}{3} \pi} 
= e^{- i \frac{2}{3} \pi} \hk , \hk z^3 = 1 \right\}$
and two vortices with $z (k = 1) = e^{\frac{2i}{3} \pi}$ can fuse to a vortex
with $z (k = 2) = z (k = 1)^2$.
\smallskip

\no
Assume now, we perform an ``inverse blocking'', successively replacing coarser
by finer lattices. As the lattice spacing is taken to zero, these $D - 1$
dimensional hypersurfaces $\Sigma$ 
become infinitely thin. Therefore, the continuum
analogue of ideal center vortices consists in specifying $D - 1$ dimensional
hypersurfaces, which when intersected by a Wilson loop contribute a center
element to the latter. Hence, in the continuum an explicit gauge field
representation of an ideal center vortex configuration in $D$ space-time
dimensions is given by 
\be
\label{8}
\cA_\mu (\Sigma, x) = E \il_\Sigma d^{D - 1} \tilde{\sigma}_{\mu} \delta^D \lk x - \bar{x}
(\sigma) \rk \hk ,
\ee
where $\bar{x}_\mu (\sigma) = \bar{x}_\mu 
\lk \sigma_1, \dots \sigma_{D - 1} \rk$
denotes a parametrization of the $D-1$ dimensional hypersurface $\Sigma$ 
and $d^{D - 1} \tilde{\sigma}_\mu$
is the dual of the $D - 1$ dimensional volume element. 
Furthermore $E$ denotes a co-weight vector, which lives in the Cartan subalgebra
$E = E_{a_0} T_{a_0}$ and is defined by
\be
\label{11}
E^{- E (k)} = Z (k) \in Z (N) \hk , \hk k = 1, \dots, N - 1 \hk 
\ee
where the $Z (k)$ denote the center elements of  the gauge group $SU (N)$ 
given by the N'th roots of unity. (Thick analogs of such configurations have
been considered in ref. \cite{Rx}.)
\smallskip

\no
Calculating the Wilson loop for the gauge field (\ref{8}) one obtains
\be
\label{12}
e^{- \oint\limits_C \cA (\Sigma)} = e^{- E I (C, \Sigma)} = Z^{I (C, \Sigma)}
\hk ,
\ee
where
\be
\label{13}
I (C, \Sigma) = \oint\limits_C d x_\mu \oint\limits_\Sigma d^{D - 1}
\tilde{\sigma}_\mu \delta^D \lk x - \bar{x} (\sigma) \rk
\ee
denotes the intersection number between the loop $C$ and the hypersurface 
$\Sigma$. To obtain the
last relation, we have used eq. (\ref{11}). Eq. (\ref{12}) can be used as gauge
invariant definition of center vortices. In the following we will refer to the
gauge field defined in eq. (\ref{8}) as ideal center vortex. One should
emphasize, that the precise position of the open hypersurface $\Sigma$ is
irrelevant for the value of the Wilson loop. This is because the intersection
number $I (C, \Sigma)$ (\ref{13}) equals the linking number $L (C, \partial
\Sigma)$
between the loop $C$ and the boundary $\partial \Sigma$, which represents the
position of the magnetic flux of the vortex. In fact, the open hypersurface
$\Sigma$ can be deformed arbitrarily by (singular) gauge transformations
keeping,
however, its boundary $\partial \Sigma$ fixed. 
\smallskip

\no
Whether the flux of  a center vortex is electric or magnetic depends on the
position of the $D - 2$ dimensional vortex surfaces
$\partial \Sigma$ in $D$-dimensional space. For example in $D = 4$
the center vortex defined by the boundary of a purely spatial 3-dimensional
volume $\Sigma$ carries only electric flux, which is directed normal to vortex
surface $\partial \Sigma$. On the other hand a vortex $\partial \Sigma$ defined
by a hypersurface $\Sigma$ evolving in time represents at fixed time, a closed
loop and carries magnetic flux, which is tangential to vortex loop, see fig. 2.
\begin{figure}
\centerline{
\epsfysize=3.5cm
\epsffile{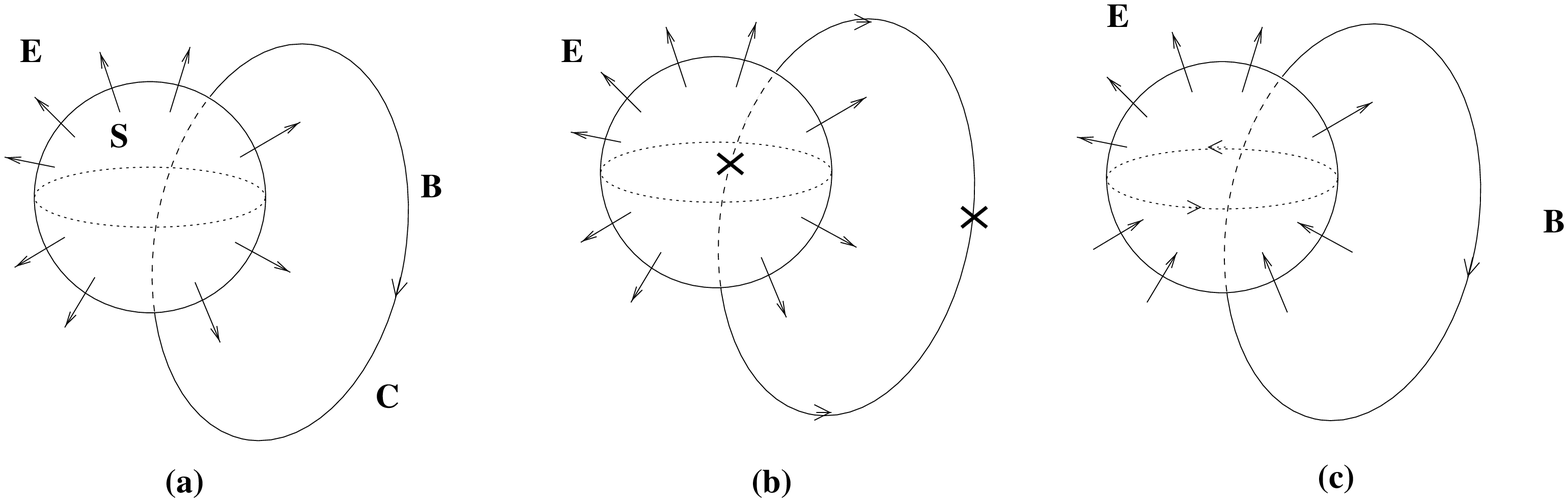}
}
\caption{Three-dimensional slice of two intersecting vortex surfaces in four
space-time dimensions. (a) globally oriented vortex surfaces, (b) the vortex
loop evolving in the fourth dimension consists of two oppositely oriented
patches connected by magnetic monopoles, (c) the vortex sphere consists of two
oppositely oriented hemispheres connected by a monopole loop. (see text)}
\label{fig2}
\end{figure}
\smallskip

\no
Consider the gauge transformation
$V (\Sigma, x ) = exp \lk - E \Omega (\Sigma, x) \rk $
defined by the solid angle $\Omega (\Sigma, x)$ 
subtended by the $D - 1$ dimensional hypersurface 
$\Sigma$ from the point $x$.
When $x$ crosses
the hypersurface $\Sigma$, the solid angle $\Omega (\Sigma, x)$ 
changes by an integer and accordingly
the gauge function $V$ changes by a center element $V \to Z V$,
but $V$ is smooth otherwise. In particular, one should note that the derivative
of $V$ along a path normal to $\Sigma$ is the same before and after the
discontinuity at $\Sigma$. Indeed ,
one can show that 
\be
V (\Sigma) \partial V^\dagger (\Sigma) = \cA (\Sigma) - a (\partial \Sigma) \hk
,
\ee
where
\be
\label{21}
a_\mu (\partial \Sigma, x) = - E \il_{\partial \Sigma} d^{D - 2} 
\tilde{\sigma}_{\mu
\nu} \partial^{\bar{x}}_\nu D \lk x - \bar{x} (\sigma) \rk
\ee
represents a vortex which carries the same flux (located on 
$\partial \Sigma$)
 as $\cA
(\Sigma, x)$. Indeed for any closed loop $C$ one has
\be
\oint\limits_C d x_\mu a_\mu (\partial \Sigma, x) = E L (C, \partial \Sigma) \hk 
\ee
with
$L (C, \partial \Sigma)$ 
being the linking number between the loop $C$ and the boundary $\partial
\Sigma$. Any Wilson loop is the same for the ideal center vortex $\cA
(\Sigma, x)$ and $a (\partial \Sigma, x)$, which is referred to in the following
as thin center vortex. One can also show that the thin center vortex $a (\partial
\Sigma, x)$ is the transversal part of the ideal center vortex 
\be
a_\mu (\partial \Sigma, x) = P_{\mu \nu } \cA_\nu (\Sigma, x) \hk , \hk P_{\mu
\nu} = \delta_{\mu \nu} - \frac{\partial_\mu \partial_\nu}{\partial^2} \hk . 
\ee
Obviously, the thin center vortex eq. (\ref{21}) manifestly depends only on 
the boundary $\partial \Sigma $,
where the flux associated with the vortex is located. 
\smallskip

\no
As an illustration
consider the magnetic flux of an infinitely thin and long solenoid, which
represents a vortex  if its magnetic flux is appropriately quantized. 
If we put the vortex on the $z$-axis the corresponding thin
vortex gauge potential is given by
\be
a (\partial \Sigma, x) = E \frac{1}{\rho} \vec{e}_\varphi \hk , \hk \rho =
\sqrt{x^2 + y^2} \hk , \hk \varphi = \arctan \frac{y}{x} \hk .
\ee
A related ideal vortex gauge potential is, for example, given by
\be
\cA (\Sigma, x) = E \vec{e}_y \delta (y) \Theta (x) \hk ,
\ee
which has support only on the positive $x$-axis, 
but yields the same value for
Wilson loops as the thin vortex.
\smallskip

\no
We are now in a position to present the continuum analogue of the maximal center
gauge. If one represents the link variables $U_\mu (x)$ in the standard fashion
by a gauge potential
$U_\mu (x) = \exp \lk - a A_\mu (x) \rk$
and takes the naive continuum limit $a \to 0$ the maximal center gauge condition
reduces to 
$- \int tr A^2 \to \min$,
which results in the Lorentz gauge
$\partial A = 0 \hk .$
This result is, however, wrong, since it relies on the expansion of the link
variables in powers of the lattice spacing around unity
$U_\mu (x) = 1 - a A_\mu (x) + \dots$,
which is not justified if the link is close to a non-trivial center element
(for example, $U_\mu = - 1$ for $SU (2)$). Even if the lattice spacing $a$ is
made smaller and smaller, during the process of gauge fixing, some of the links,
initially close to unity, are rotated to group elements close to a non-trivial
center element, for which the expansion around $U_\mu = 1$ fails. 
\bi

\no
A careful
analysis \cite{R6} shows that the continuum analogue of the maximal center 
gauge condition is given by
\be
- \min\limits_{\partial \Sigma} \min\limits_g \int tr \lk A^g - a (\partial
\Sigma) \rk^2 \hk ,
\ee
where the minimization has to be performed with respect to all continuum gauge
transformations $g$ and to all vortex sheets $\partial \Sigma$. As the result of
this double minimization procedure, a given gauge potential $A$ is brought
as close
as possible to a thin center vortex $a (\partial \Sigma)$. In this sense the
maximal center gauge in the continuum theory represents a ``maximal vortex
gauge''. 
\bi

\no
For a fixed vortex
sheet $\partial \Sigma$ minimization with respect to the gauge group yields the
condition 
\be
\left[ \partial + a (\partial \Sigma), A \right] = 0 \hk ,
\ee
which is the background gauge fixing with the thin vortex $a (\partial \Sigma)$
figuring as background field.

\section{Topology of center vortices}

\bi

\no
There is an essential difference between the center vortices on the lattice and
in the continuum theory. After center projection on the lattice, the direction
of the magnetic flux of the vortex is lost, while in the continuum theory the
center vortices are given by oriented (patches of) surfaces, where the flux
direction is defined by the orientation of the surfaces. The orientation of the
vortices, however, is crucial for their topological properties as will be shown
later. 
\bi

\no
A closed vortex surface $\partial \Sigma$ need not be globally oriented
but can consist of several oriented patches $S_k$, 
$\partial \Sigma = \bigcup\limits_k S_k $ .
Furthermore, these patches will in general carry flux corresponding to different
non-trivial center elements, that is to different co-weight vectors $E$. We will
now see that the boundaries between different oriented vortex patches host
magnetic monopole loop currents. Let us first illustrate this in $D = 3$.
\bi

\no
Consider an ordinary magnetic monopole and an anti-monopole, which are
connected by a Dirac string, see fig. 3.
\begin{figure}
\centerline{
\epsfysize=3cm
\epsffile{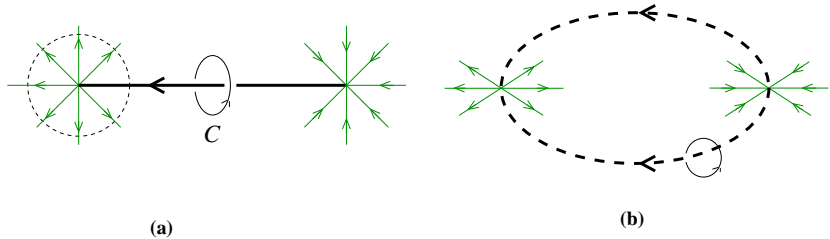}
}
\caption{Illustration of the connection between Dirac string (a) and center
vortex (b) (see text)}
\label{fig3}
\end{figure}

 Since the Dirac string is a gauge
artifact, it has to be invisible for any Wilson loop, i.e. contribute
unity to the Wilson loop, just like a vanishing gauge potential. This must hold,
in particular, for loops non-trivially linked to the Dirac string, which
requires the magnetic flux contained in the Dirac string to be quantized in
multiples of $2 \pi i$
\be
\phi = \il_S d \vec{\sigma} \vec{B} = \oint\limits_{\partial S = C} d \vec{x}
\vec{A} = 2 \pi i \hk ,
\ee
such that $e^ {- \phi} = 1$ .
This is the usual Dirac quantization condition. In fact
 the magnetic flux through
a sphere $S_2$ around the monopole position (excluding from $S_2$ the piercing
point of the Dirac string) is the same as the flux of the Dirac string flowing
into the monopole. Hence quantization of the flux of the Dirac string implies
 the quantization of the magnetic charge of the
monopole. 
\bi

\no
Assume now that the
magnetic flux of the Dirac string is split into two equal portions, see fig. 3 .
The half Dirac string carries the flux $\phi/2 = \pi$ and hence
contributes $e^{- i \pi } = (- 1)$ 
to a Wilson loop non-trivially linked to it. This is the
``largest'' contribution a field configuration can contribute to the Wilson
loop (in the sense that $-1$ is the phase which has the largest distance from
the trivial phase $1$). The half Dirac string, obviously,
represents a center vortex. By splitting the Dirac string into two strings of
half flux, we have created a center vortex. This center vortex consists
of two strings, which are connected by magnetic monopoles. The crucial point is,
that at the monopole position, the direction of the magnetic flux (that is the
orientation of the vortex) is reversed. This feature will survive in $D = 4$,
where vortices are closed sheets and magnetic monopoles move on closed loops.
\bi

\no
The reversal of the orientation on the vortex sheet by magnetic monopole loops
can also be seen in an alternative way: The vortex line (sheet) with reversed
orientations can be also interpreted as an oriented vortex covered with an
oppositely directed (open) Dirac string (sheet). This is because the Dirac
string (sheet) carries twice the flux of a center vortex. The boundary of the 
open Dirac string world sheet represents the word line of a Dirac magnetic 
monopole (that is having a magnetic charge in accord with Dirac's quantization 
condition).
\bi

\no
In this way non-oriented
closed magnetic vortex sheets consist of oriented surface
pieces joined by magnetic monopole current loops. Such a
non-oriented closed magnetic sheet defines a center vortex, which still gives
the same contribution to a Wilson loop as the corresponding oriented vortex (in
the absence of a monopole loop). Thus for the confinement properties measured by
the Wilson loop, the orientation of the vortex sheet and hence the magnetic
monopole currents are irrelevant. The Abelian magnetic monopole currents are,
however, necessary in order to generate a non-vanishing Pontryagin index for
vortex configurations as will be shown below.
\bi

\no
The gauge fields can be topologically classified by the Pontryagin index (second
Chern class)
\bea
\label{x42}
\nu [A] = - \frac{1}{16 \pi^2} \int d^4 x tr F_{\mu \nu} \tilde{F}_{\mu \nu}
 =  \frac{1}{4 \pi^2} \int d^4 x E^a_i B^a_i \hk ,
\eea
where $E^a_i = F^a_{i 0}$ and $B^a_i = \frac{1}{2} \epsilon_{i j k} F^a_{j k}$
are the corresponding colour electric and magnetic fields. From the above 
relation it is clear that topologically non-trivial center vortices with 
$\nu \neq 0$
have to carry both electric and magnetic flux. Furthermore, it is also easy to
see that magnetic monopole currents are absolutely necessary on the Abelian
center vortex sheet in order to generate a non-vanishing Pontryagin index. In
fact, using the definition of the vortex field strength in terms of the ideal
center vortex gauge potential 
$\cA_\mu (\Sigma, x)$, $\cF_{\mu \nu} (\partial \Sigma) = 
\partial_\mu \cA_\nu (\Sigma) - \partial_\nu \cA_\mu (\Sigma)$ 
and performing a partial integration, one obtains
\be
\nu  = - \frac{1}{8 \pi^2} \il_M \partial_\mu tr \lk \cA_\nu \tilde{\cF}_{\mu
\nu} \rk + \frac{1}{8 \pi^2} \il_M tr \lk \cA_\nu \partial_\mu
\tilde{\cF}_{\mu \nu} (\partial \Sigma) \rk \hk .
\ee
Using Gauss' law in the first term and the definition of the monopole current,
$j^m_\mu = \partial_\nu \tilde{\cF}_{\mu \nu}$,
in the second one, we obtain
\bea
\nu & = &  - \frac{1}{8 \pi^2} \il_{\partial \cM} d \sigma_\mu tr \lk \cA_\nu
\cF_{\mu \nu} \rk 
+ \frac{1}{8 \pi^2} \il_\cM tr \lk \cA_\nu j^m_\nu \rk \nonumber\\
& \equiv & \nu^{(1)} + \nu^ {(2)} \hk .
\eea
The surface term $\nu^{(1)}$ vanishes unless the integrand contains
singularities, which give rise to internal surfaces wrapping the
singularities. Such internal surfaces precisely arise in the presence of
magnetic charges, which may be magnetic monopoles or more extended magnetic
charge distributions such as line or surface charges. The second term can be
cast into the form $\nu^{(2)} = - \frac{1}{4} L (\partial \Sigma, C)$
where $L (\partial \Sigma, C)$ is the linking number
between the vortex
surface $\partial \Sigma$ and the monopole loop $C$. This shows that a
non-zero Pontryagin index indeed requires the existence of magnetic charges in
the Abelian projected configurations, to which the center vortices introduced
above belong.
\bi

\no 
An alternative expression for the Pontryagin index of center vortex sheets,
which does not make explicit reference to magnetic monopoles, can be obtained by
inserting the explicit expression
for the field strength of a center vortex, 
$\cF_{\mu \nu} (S) = E \il_S d^2 \sigma_{\mu \nu} \delta^4 \lk x - \bar{x} (\sigma)
\rk$ into eq. (\ref{x42}). 
Using (for
$SU (2)$) $tr (E E) = - 2 \pi^2$, one finds immediately 
\be
\nu \left[ \cA (\Sigma) \right] = \nu \left[ a (\partial \Sigma) \right] =
\frac{1}{4} I (\partial \Sigma, \partial \Sigma) \hk ,
\label{nuisn}
\ee
where
\be
\label{44}
I (S_1, S_2) = \frac{1}{2} \il_{S_1} d^2 \tilde{\sigma}_{\mu \nu} \int d^2
\sigma'_{\mu \nu} \delta^4 \lk \bar{x} (\sigma) - \bar{x} (\sigma') \rk
\ee
is the oriented intersection number between two 2-dimensional (in general open)
surfaces $S_1, S_2$ in $R^4$. From its definition, it follows that 
(in $ D = 4$) $I (S_2, S_1) = I (S_1, S_2)$. Generically two 2-dimensional 
surfaces intersect in
$R^4$ at isolated points. Obviously the (self-)intersection number $I \lk
\partial \Sigma, \partial \Sigma \rk$ receives contributions only from those 
points $\bar{x} (\sigma) = \bar{x} (\sigma^{\prime } )$ where the intersecting 
surface patches give rise to 
four linearly independent tangent vectors (otherwise $d \sigma_{\mu \nu} d
\tilde{\sigma}_{\mu \nu} = 0$). Such points will be referred to as singular
points. One can distinguish two principally different types of singular
points \cite{R6,preptop}:

\begin{enumerate}
\item transversal intersection points, for which
$x (\sigma) = \bar{x} (\sigma') \hk {\rm and} \hk \sigma \neq \sigma'$
\item twisting (or writhing) points
$\bar{x} (\sigma) = \bar{x} (\sigma') \hk {\rm and} \hk \sigma = \sigma'$
\end{enumerate}
\begin{figure}[h]
\centerline{
\epsfysize=5cm
\epsffile{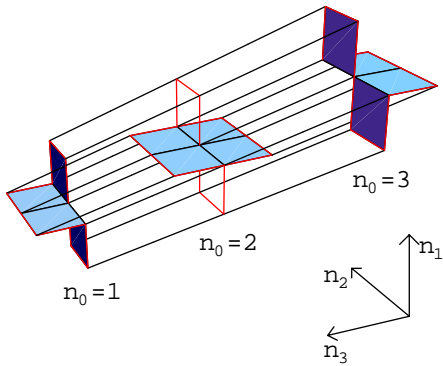}
}
\caption{Sample vortex surface configuration, composed of elementary
squares on a hypercubic lattice. At each (lattice) time
$n_0 $, shaded plaquettes are part of the vortex surface. These
plaquettes are furthermore connected to plaquettes running in time
direction; their location can be inferred most easily by keeping in
mind that each link of the configuration is connected to exactly
two plaquettes (i.e. the surface is closed and contains no intersection
lines). Note that the two non-shaded plaquettes at $n_0 =2$ are {\em not}
part of the vortex; only the two sets of three links bounding them are.
These are slices at $n_0 =2$ of surface segments running in time
direction from $n_0 =1$ through to $n_0 =3$. Sliced at $n_0 =2$, these
surface segments show up as lines. Furthermore, by successively assigning
orientations to all plaquettes, one can convince oneself that the
configuration is orientable.}
\label{fig5}
\end{figure}
Examples of the singular points of a lattice realization of a vortex sheet
 are 
given in figure \ref{fig5} \cite{preptop}. 
The surface in figure \ref{fig5} is closed, 
orientable, has one genuine self-intersection point (precisely at the 
center of the configuration). 
Additionally, it has twisting (or writhing) points, at $n_0 =1$ and $n_0 =3$ 
in Fig. \ref{fig5}, as well as at $n_0 =2$ at the front and back edges of 
the configuration (from the viewer's perspective).
\bi

\no
Transversal intersection points yield a contribution $\pm 1$ to the (oriented)
intersection number, where the sign depends on the relative orientation of the
two 
intersecting pieces\footnote{Later on, 
it will be seen that each transversal intersection 
point actually contributes twice to the self-intersection
number, so that these points contribute $\pm 2$ to 
$I (\partial \Sigma, \partial \Sigma)$.}. (This is easily seen by considering the intersection of two
orthogonal intersecting planes.) Twisting (or writhing) points yield positive 
or negative contributions of modulus smaller than one to the self-intersection
number.
\bi

\no
It is well-known in topology that the self-intersection number of closed
2-dimensional surfaces in $R^4$ is a multiple of 4, so that the Pontryagin index
(\ref{nuisn}) is indeed integer valued.
It is also known that the self-intersection number of closed globally oriented
2-dimensional surfaces in $R^4$ vanishes. This implies that the Pontryagin 
index vanishes for
globally oriented vortex surfaces. Consequently non-orientedness of the vortex
surfaces is crucial for generating a non-vanishing Pontryagin index. To
illustrate this, consider the following 3-dimensional slice of two intersecting
$SU (2)$ vortex surfaces in $D = 4$ dimensions, cf. fig. 2. Let the one
vortex be located entirely within the spatial 
3-dimensional slice of our 4-dimensional
space-time manifold. It is therefore visible as a closed surface $S$, namely the
sphere
in fig. 2. On the other hand, let the other vortex extend into the time
dimension not displayed in fig. 2. At a fixed time, it 
is then visible in 3-space as a closed loop $C$.
Let $C$ intersect $S$ at two points as shown in fig. 2. 
If $S$ and $C$ are both globally oriented as in fig. 2 a, the
(oriented) intersection number (in $D = 3$)
vanishes, since the two intersection points occur with opposite relative
orientation between the intersecting manifolds. Thus, a non-zero intersection
number requires at least one of the two intersecting closed manifolds to be
not globally oriented. Globally non-oriented surfaces consist nevertheless of
oriented patches. Assume for example the sphere $S$ in fig. 2 c to consist
of two hemispheres of opposite orientation connected by a monopole loop. 
In this case the contributions to the oriented intersection number from
the two intersection points no longer cancel but add, giving an
intersection number of two.
\bi

\no
It should come as no surprise that the magnetic monopoles are required to
endow the vortex sheets with a non-trivial topological structure. This is
because in certain Abelian gauges, the Pontryagin index can be expressed
entirely in terms of magnetic charges. Let us consider for example the
Polyakov gauge, where one diagonalizes the Polyakov line 
\be
\Omega (\vec{x}) = \exp
\lk - \il d x_0 A_0 (x) \rk = V^\dagger (x) \omega (x) V (x) \to \omega
(x) \hk
\ee
by performing the gauge transformation $A \to A^V = V a V^\dagger + V
\partial V^\dagger$ with the coset matrix $V \in SU (2) / U (1)$. At those
isolated points in space $\vec{x}_i$ where the Polyakov line becomes a center
element $\Omega (\vec{x}_i ) = (- 1)^{n_i} \in Z (2)$, the inhomogeneous part of
the gauge transformed field $V \partial V^\dagger$ develops a magnetic monopole,
whose charge is topologically quantized and given by the winding number $m [V]
\in \pi_2 \lk SU (2) / U (1) \rk$ of the mapping $V (\vec{x})$ from 
$S_2$ around the magnetic monopole at $\vec{x}_i$ 
into the coset $SU (2) / U (1) \sim S_2$. The
Pontryagin index is then given by the exact relation
\be
\nu = \sli_i n_i m_i \hk ,
\ee 
where the sum runs over all
magnetic monopoles.
This relation was first derived in 
ref. \cite{R7} and was later rederived in
\cite{R8}. 
\bi

\no
\section{Concluding remarks}
\bi

\no
The topological properties of center vortices discussed in this talk are
directly relevant for an understanding of the anomalous generation of the 
$\eta'$ mass within the vortex picture of the QCD vacuum, since this mass
is determined by the topological susceptibility through the Witten-Veneziano 
formula. Furthermore, these topological properties are likely to play a
role in the spontaneous breaking of chiral symmetry, in analogy to the
arguments advanced in the framework of instanton models. To
explain these phenomena, it is necessary to study the occurence of the various
intersection points within the vortex ensemble \cite{preptop}. Since these 
intersection points carry non-trivial topological charge,
by the Atiyah-Singer index theorem, they
should give rise to fermionic zero modes, which within the instanton picture of
the QCD vacuum have proven responsible for the spontanous breaking of chiral
symmetry. In view of the lattice result, which shows that the quark condensate
disappears when the center vortices are removed \cite{R5}, 
it is our strong belief that
the vortex picture cannot only give an appealing picture of confinement and the
deconfinement phase transition, but at the same time can provide an
understanding of spontaneous breaking of chiral symmetry and the emergence of
the topological susceptibility.
\bi

\no
\section*{Acknowledgments}
R.~Bertle and M.~Faber are gratefully acknowledged for providing the 
MATHEMATICA routine with which the image in Fig. \ref{fig5} was generated.
\bi

\no


\section*{References}
\begin{thebibliography}{99}
\bibitem{R1}L. Del Debbio, M. Faber, J. Greensite and \v{S}. Olejnik, Phys. Rev.
{\bf D55} (1997) 2298

\bibitem{R2}K. Langfeld, H. Reinhardt and O. Tennert, Phys. Lett. {\bf B419}
(1998) 317\\
M. Engelhardt, K. Langfeld, H. Reinhardt, O. Tennert, Phys. Lett. {\bf B431}
(1998) 141

\bibitem{R3}K. Langfeld, O. Tennert, M. Engelhardt, H. Reinhardt, Phys. Lett.
{\bf B452} (1999) 301\\
M. Engelhardt, K. Langfeld, H. Reinhardt, O. Tennert, Phys. Rev. {\bf D61}
(2000) 054504

\bibitem{selprep} M. Engelhardt, H. Reinhardt, hep-lat/9912003

\bibitem{R4}T. Kovacs and E. Tomboulis, hep-lat/0002004

\bibitem{R5} Ph. de Forcrand, M. D'Elia, Phys. Rev. Lett. {\bf 82} (1999) 4582
\bibitem{R6} M. Engelhardt, H. Reinhardt, Nucl. Phys. {\bf B567} (2000) 249
\bibitem{preptop} M. Engelhardt, hep-lat/0004013
\bibitem{R7} H. Reinhardt, Nucl. Phys. {\bf B503} (1997) 505
\bibitem{R8} C. Ford, U.G. Mitreuter, T. Tok, A. Wipf, J.M. Pawlowski, Ann.
Phys. {\bf 269} (1998) 26\\
O. Jahn, F. Lenz, Phys. Rev. {\bf D58} (1998) 085006\\
M. Quandt, H. Reinhardt, A. Schaefke, Phys. Lett. {\bf B446} (1999) 290
\bibitem{Rx} J. M. Cornwall, Phys. Rev. {\bf D59} (1999) 125015
\end{thebibliography}
\end{document}